\documentclass[prl,twocolumn,a4paper]{revtex4-1}
\usepackage{graphicx}
\usepackage{enumitem}
\usepackage{amsmath}

\usepackage{amsfonts}

\begin{document}

\title{Exact Thermal Distributions in Integrable Classical and Quantum Gases}

\author{Manuel Valiente}
\affiliation{Departamento de F{\'i}sica, CIOyN, Universidad de Murcia, 30071 Murcia, Spain }

\begin{abstract}
We consider one-dimensional, integrable many-body classical and quantum systems in thermal equilibrium. In the classical case, we use the classical limit of the Bethe equations to obtain a self-consistent integral equation whose solution gives the distribution of asymptotic Bethe momenta, or rapidities, as well as the classical partition function in the canonical ensemble, and the thermal energy dispersion. For quantum gases, we obtain a similar integral equation, albeit in the grand canonical ensemble, with completely analogous results. We apply our theory to the classical and quantum Tonks and Calogero-Sutherland models, and our results are in perfect agreement with standard calculations using Yang-Yang thermodynamics. Remarkably, we show in a straightforward manner that the thermodynamics of the quantum Calogero-Sutherland model is in one-to-one correspondence with the ideal Fermi gas upon simple rescalings of chemical potential and density.
\end{abstract}


\maketitle

\paragraph{Introduction.} Non-relativistic integrable many-body systems constitute excellent playgrounds for testing non-perturbative phenomena. These systems are often exactly solvable \cite{Calogero2001,Sutherland2004}. Among the most well-known techinques for their solution, the Bethe ansatz is a particularly powerful method in quantum integrable models \cite{Sutherland2004,Karbach1997a}. For classical integrable gases, the techique of Lax pairs \cite{Calogero2001,Hitchin1999}-- which may also be adapted to the quantum case \cite{Sutherland2004,Moser1975,Calogero1975a,Calogero1975b}-- allows the construction of $N$ independent constants of motion for $N$ particles that completely characterize the Newtonian dynamics of the many-body system. Isolated integrable models do not thermalize in the usual sense \cite{Rigol2008,Kollath2007,Manmana2007}, simply because they feature a thermodynamically large number of conserved quantities, only one of them being the Hamiltonian. If an integrable system couples to external sources, then it may thermalize as in standard statistical mechanics \cite{Huang1987}. However, it is remarkably simpler to calculate exact thermodynamic properties in the quantum case -- using the thermodynamic Bethe ansatz (also called Yang-Yang thermodynamics \cite{Yang1969}) -- than in the classical case where, in general, the partition functions had to be calculated in the standard manner so far. 

In this Letter, we study classical and quantum mechanical integrable many-body systems in thermal equilibrium. In the classical case, we provide an exact method to calculate the classical partition function in the canonical ensemble by considering the classical limit of the Bethe ansatz equations. As a byproduct, the thermodynamic energy dispersion and the distribution of asymptotic (or Bethe) momenta, also known as rapidities, are also calculated. Releasing the system after thermalization -- be it usual or generalized \cite{Rigol2008,Calabrese2016,Vidmar2016}-- the momentum distribution, once the system becomes dilute enough, coincides with the rapidity distribution . This has been measured in a number of ground breaking experiments with isolated atomic gases \cite{Malvania2021,Wilson2020,Li2023} (in the quantum regime). For quantum gases, we work in the grand canonical ensemble, and obtain all thermodynamic properties via an integral equation, which is completely analogous to the classical case, fully circumventing the previous requirement to calculate the thermodynamic energy dispersions, which is the starting point in Yang-Yang thermodynamics \cite{Sutherland2004,Yang1969}. In both the classical and quantum cases, we study the Tonks \cite{Tonks1936} (Tonks-Girardeau \cite{Girardeau1960}) and Calogero-Sutherland \cite{Sutherland1971a,Sutherland1971b} gases and prove that the thermodynamics of the Calogero-Sutherland model can be reduced to that of a non-interacting Fermi gas with rescaled chemical potential and density.

\paragraph{Asymptotic Bethe ansatz.} We consider spinless bosons or fermions. Our starting point is the asymptotic Bethe ansatz \cite{Sutherland2004,Karbach1997a}, which represents the long-distance behaviour of the many-body wave function for an integrable system. For bosons, this reads, in the limit $|x_i-x_j|\to\infty$ for all $i\ne j$,
\begin{equation}
    \psi=\sum_P\exp\left[i\sum_{j=1}^Nk_{P_j}x_j+i\sum_{i<j}\tilde{\theta}_{P_i,P_j}/2\right],
\end{equation}
where the sum runs over all permutations of asymptotic momenta $\{k_1,k_2,\ldots,k_N\}$, and where $\tilde{\theta}_{P_i,P_j}$ is the two-body phase shift. To avoid confusions with different notational conventions, note that for two bosons, in the relative coordinate $x=x_1-x_2$, the two-body asymptotic wave function for bosons takes the form $\psi(x)=\sin(k|x|+\tilde{\theta}(k)/2)$, with $k=(k_1-k_2)/2$ the relative asymptotic momentum. For fermions, it is given by $\phi=\prod_{i<j}\mathrm{sgn}(x_i-x_j)\psi$. We define, following standard conventions for the Bethe ansatz \cite{Sutherland2004}, $\theta(k_i-k_j)\equiv \tilde{\theta}((k_i-k_j)/2)$. For classical particles, the relevant quantities are $\lambda_{P_j}=\hbar k_{P_j}$ as $\hbar\to 0$, and the classical phase shifts $\varphi(\lambda_{P_i}-\lambda_{P_j})=\lim_{\hbar\to 0}\hbar \theta(k_{P_i}-k_{P_j})$. Note, however, that the classical phase shifts can be calculated directly via the WKB approximation \cite{Sakurai}, and a full quantum solution is not required.

\paragraph{Classical gases.} The classical analog to the set of Bethe equations for the Bethe momenta $\{\lambda_j\}_{j=1}^{N}$, given a set $\{p_j\}_{j=1}^N$ of momentum-valued quantities, is given by
\begin{equation}
    p_j=\lambda_j+\frac{1}{L}\sum_{\ell (\ne j)=1}^N\varphi(\lambda_j-\lambda_{\ell}),\label{eq:Bethe1}
\end{equation}
where $\varphi(\lambda)$ is the classical phase shift. In the thermodynamic limit, if the distribution density of Bethe momenta is $\Pi(\lambda)$ (with units of inverse momentum), we can make the replacement
    $\sum_{\ell=1}^Nf(\lambda_{\ell})\to N \int d\lambda \Pi(\lambda)f(\lambda)$.
The Bethe equations (\ref{eq:Bethe1}) for the variable $p$ take the form
\begin{equation}
    p=\lambda+\rho\int d\gamma \Pi(\gamma) \varphi(\lambda-\gamma).\label{eq:Bethe2}
\end{equation}
Now we use the obvious fact that, in thermodynamic equilibrium, at temperature $T$, the distribution $\pi(p)$ of $p$'s is a Boltzmann distribution, that is,
\begin{equation}
    \pi(p)=\frac{e^{-\beta\epsilon(p)}}{Z_0(\beta)},
\end{equation}
where $\beta=1/kT$, and $\epsilon(p)$ is a suitable even function of $p$ called the thermodynamic dispersion \cite{Sutherland2004}, satisfying
\begin{equation}
    \epsilon(p(\lambda))=\frac{\lambda^2}{2m}.\label{eq:classicaldispersion}
\end{equation}
The ($N$-th root of) the canonical partition function is then given by $Z_0(\beta)$, with
\begin{equation}
    Z_0(\beta)=\int dpe^{-\beta \epsilon(p)}.
\end{equation}
Here, we find relations that completely circumvent the need to calculate the dispersion $\epsilon(p)$ as a previous step, and which, moreover, allow for a simple calculation of the distribution of Bethe momenta, as well as the partition function (and derive all the thermodynamics of the system from it).

The distributions $\Pi(\lambda)$ and $\pi(p)$ are related via the transformation
\begin{equation}
    \Pi(\lambda)=p'(\lambda)\pi(p(\lambda)).\label{eq:trans}
\end{equation}
Differentiating Eq.~(\ref{eq:Bethe2}) with respect to $\lambda$, and inserting the resulting expression into Eq.~(\ref{eq:trans}), we obtain
\begin{equation}
    \Pi(\lambda)=\frac{e^{-\beta\lambda^2/2m}}{Z_0(\beta)}\left[1+\rho\int d\gamma \varphi'(\lambda-\gamma)\Pi(\gamma)\right].\label{eq:inteq}
\end{equation}
Solving Eq.~(\ref{eq:inteq}), with fixed $Z_0(\beta)$, gives a solution $\Pi(\lambda)\equiv F(\lambda;Z_0(\beta))$. We obtain the partition function self-consistently by imposing normalization of the distribution $\Pi(\lambda)$, via
\begin{equation}
\int d\lambda F(\lambda;Z_0(\beta))=1.\label{eq:normalization}
\end{equation}
The simplest way to obtain the dispersion $\epsilon(p)$, once $\Pi(\lambda)$ is known, consists of solving Eq.~(\ref{eq:Bethe2}) for $\lambda=\lambda(p)$, and obtain $\epsilon(p)$ from Eq.~(\ref{eq:classicaldispersion}).

After characterizing the zero-temperature physics, we will study the two simplest integrable models, namely Calogero-Sutherland and Tonks gases. 

\paragraph{Zero-temperature limit.} The lowest energy configuration, for classical particles, is obtained by setting $p_j=0$ for all $j$, and setting $\Pi(\lambda)=G(\lambda)\theta(\lambda_*-|\lambda|)$, where $\lambda_*$ ($>0$) is a momentum scale such that 
\begin{equation}
    \int_{-\lambda_*}^{\lambda_*}d\lambda G(\lambda)=1.\label{eq:normalization0T}
\end{equation}
Setting $p=0$, and differentiating Eq.~(\ref{eq:Bethe2}) with respect to $\lambda$, we have
\begin{equation}
    \rho\int_{-\lambda_*}^{\lambda_*}d\gamma G(\gamma)\varphi'(\lambda-\gamma)=-1.\label{eq:inteq0T}
\end{equation}

\paragraph{Tonks gas.} The Tonks gas consists of $N$ particles that interact via a hard rod potential of diameter $a$. The phase shifts are simply $\varphi(\lambda)=-\lambda a$. For this model, we know that, given an initial distribution of canonical momenta $\Lambda_P(0)=\{P_j(0)\}_{j=1}^{N}$, after Newtonian evolution, the set $\Lambda_{P}(t)=\{P_j(t)\}_{j=1}^N$ is invariant, since only a reordering of canonical momenta occurs. Therefore, identifying the set (disregarding order) $\Lambda_{P}(0)$ with the set of Bethe momenta $\Lambda_{\lambda}=\{\lambda_j\}_{j=1}^N$, it is obvious that the thermal distribution of Bethe momenta in the thermodynamic limit is simply the Boltzmann distribution
\begin{equation}
    \Pi(\lambda)=\sqrt{\frac{\beta}{2\pi m}}e^{-\beta\lambda^2/2m}.\label{eq:tonksboltzmann}
\end{equation}
Inserting the above relation and the Tonks phase shifts into Eq.~(\ref{eq:inteq}), the partition function $Z_0(\beta)$ takes the form
\begin{equation}
    Z_0(\beta)=(1-\rho a)\sqrt{\frac{ 2\pi m}{\beta}},\label{eq:tonkspartition} 
\end{equation}
which is exactly what is obtained performing a direct calculation of the partition function in the thermodynamic limit \cite{Tonks1936}. Of course, knowing Eq.~(\ref{eq:tonksboltzmann}) beforehand is not necessary. The integral equation~(\ref{eq:inteq}) is easily solved, since $\Pi(\lambda)\exp(\beta\lambda^2/2m)$ is a constant ($\varphi'(\lambda)=-a$), and imposing Eq.~(\ref{eq:normalization}), we immediately obtain Eq.~(\ref{eq:tonkspartition}).

The classical dispersion is straightforwad to obtain, since $\lambda=p/(1-\rho a)$ at all temperatures, so 
\begin{equation}
    \epsilon(p)=\frac{p^2}{2(1-\rho a)^2m}.
\end{equation}

At zero temperature, Eq.~(\ref{eq:inteq0T}), under condition (\ref{eq:normalization0T}) is immediately solved by 
\begin{equation}
    G(\lambda)=\delta(\lambda),
\end{equation}
where $\delta(\lambda)$ is the Dirac delta function, while $\lambda_*$ is arbitrary and positive. This, of course, was expected, since the initial set of momenta does not change in time. It is also obtained as the limit of $\beta\to \infty$ in the Boltzmann distribution, Eq.~(\ref{eq:tonksboltzmann}).

\paragraph{Calogero-Sutherland model.} For the inverse square interaction, the phase shifts are given by $\varphi(\lambda)=-(c/2)\mathrm{sgn}(\lambda)$, where $c$ is a positive constant with dimensions of action (like $\hbar$). Introducing the phase shifts into the integral equation (\ref{eq:inteq}) for $\Pi(\lambda)$, we obtain
\begin{equation}
    F(\lambda;Z_0(\beta))=\frac{e^{-\beta\lambda^2/2m}}{Z_0(\beta)+\rho ce^{-\beta\lambda^2/2m}}.\label{eq:FCalogero}
\end{equation}
Normalization of the distribution, Eq.~(\ref{eq:normalization}), gives the following transcendental equation for the partition function
\begin{equation}
    \mathrm{Li}_{1/2}\left(-\frac{\rho c}{Z_0(\beta)}\right)=-\rho c\sqrt{\frac{\beta}{2\pi m}}.
\end{equation}

\begin{figure}[t]
    \centering
    \includegraphics[width=\columnwidth]{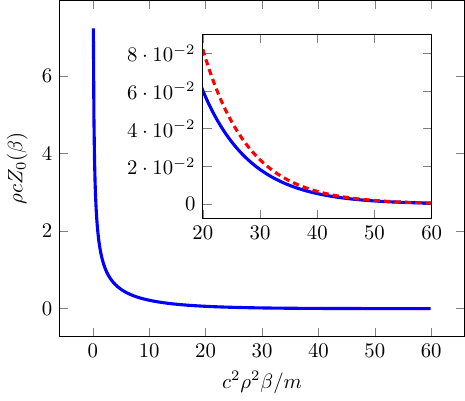}
    \caption{Blue solid line: $N$-th root, $Z_0(\beta)$ of the partition function, in the thermodynamic limit, for the classical Calogero-Sutherland model. Inset: same as in main figure; red dashed line: low-temperature limit of $Z_0(\beta)$, Eq.~(\ref{eq:lowT}).}
    \label{fig:Z0-Sutherland}
\end{figure}

\begin{figure}[t]
    \centering
    \includegraphics[width=\columnwidth]{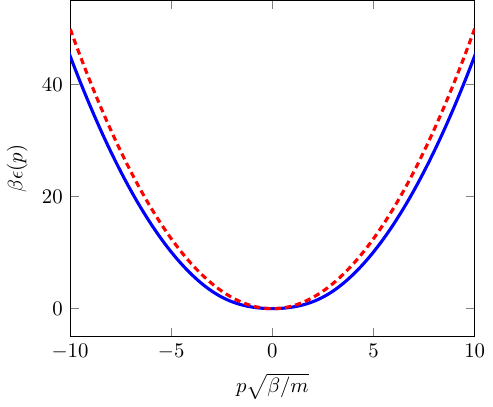}
    \caption{Blue solid line: dispersion for the classical Calogero-Sutherland model with $\rho c/\sqrt{m\beta}=1$; red dashed line: non-interacting, Galilean dispersion $p^2/2m$.}
    \label{fig:dispersion}
\end{figure}

At zero temperature, Eq.~(\ref{eq:inteq0T}) under condition (\ref{eq:normalization0T}) is solved by 
\begin{equation}
    G(\lambda)=\frac{1}{\rho c},
\end{equation}
and $\lambda_*=\rho c/2$. Since Eq.~(\ref{eq:FCalogero}) must have the correct zero-temperature limit, we impose $\lim_{\beta\to\infty}F(\lambda_*;Z_0(\beta))=1/2\rho c$, obtaining the low-temperature limit of $Z_0(\beta)$ 
\begin{equation}
    Z_0(\beta)\sim \rho c e^{-\beta (\rho c)^2/8m},\hspace{0.1cm} \beta\to \infty.\label{eq:lowT}
\end{equation}
In Fig.~\ref{fig:Z0-Sutherland}, we plot $Z_0(\beta)$ as a function of inverse temperature. As is observed in its inset, Eq.~(\ref{eq:lowT}) is in very good agreement with the exact values for sufficiently low temperatures. 
The dispersion $\epsilon(p)$, is obtained from Eqs.~(\ref{eq:Bethe2}) and (\ref{eq:classicaldispersion}), by quadrature, which reads in this case,
\begin{equation}
    p=\lambda-\frac{\rho c}{2}\int d\gamma \Pi(\gamma)\mathrm{sgn}(\lambda-\gamma),
\end{equation}
and is plotted in Fig.~\ref{fig:dispersion}, where it is seen that it is quite different from the Galilean ($p^2/2m$) dispersion.

\paragraph{Quantum gases.} The quantum case is almost equivalent to the classical situation. The main difference being that each $p_j$ in Eq.~(\ref{eq:Bethe1}) occurs at most once. Therefore, the Boltzmann distribution $\pi(p)$ is replaced here with the Fermi-Dirac distribution, and so we work in the grand canonical ensemble. It is also convenient to work with $k^{(0)}\equiv p/\hbar$, as well as $k\equiv \lambda/\hbar$. The quantum phase shifts are then $\theta(k)\equiv \varphi(\hbar k)/\hbar$. Equation (\ref{eq:Bethe1}) is valid, with the replacements just considered. In the grand canonical ensemble, the thermodynamic limit is taken by the substitution $\sum_{\ell}f(\lambda_{\ell})\to \langle N \rangle \int dk\Pi(k)f(k)$, and the density $\rho$ in Eq.~(\ref{eq:Bethe2}) is replaced with the average density $\rho(\mu)=\langle N\rangle$, with $\mu$ the chemical potential. It enters the distribution of $k^{(0)}$ as 
\begin{equation}
\pi(k^{(0)})=\frac{1}{2\pi\rho(\mu)}\frac{1}{e^{\beta (\epsilon(k^{(0)})-\mu)}+1}\equiv \frac{f(\epsilon(k^{(0)}))}{2\pi\rho(\mu)},
\end{equation}
where the factor $1/2\pi \rho(\mu)$ -- which must be calculated self-consistently -- ensures unit normalization of $\pi(k^{(0)})$.
Proceeding just as in the classical case, we obtain the following integral equation for the distribution of Bethe momenta $\Pi(k)\equiv G(k)f(\hbar^2k^2/2m)/2\pi\rho(\mu)$,
\begin{equation}
    G(k)=1+\int \frac{dk'}{2\pi}\theta'(k-k')f(\hbar^2k'^2/2m)G(k').\label{eq:PiQuantum}
\end{equation}
The self-consistency relation to complement Eq.~(\ref{eq:PiQuantum}) is given, once more, by normalization of $\Pi(k)$,  $\int dk \Pi(k) = 1$. This is nothing but an equation for the chemical potential given the density $\rho$, or for the density $\rho$ given the chemical potential. 

We do not consider the zero-temperature limit for quantum gases, since its solution is already well known \cite{Sutherland2004}. We study now the quantum versions of Tonks (called Tonks-Girardeau) and Calogero-Sutherland models.

\paragraph{Tonks-Girardeau gas.} The solution is straightforward, as $\theta(k)=-ka$ implies that
\begin{equation}
    \Pi(k)=\frac{1}{2\pi\rho(1+I(\beta,\mu)a)}f(\hbar^2k^2/2m),
\end{equation}
where 
\begin{equation}
2\pi I(\beta,\mu)=\int dk f(\hbar^2k^2/2m)=-\sqrt{\frac{\pi \beta}{2m}}\mathrm{Li}_{1/2}(-e^{\beta\mu}).
\end{equation}
The relation between density and chemical potential follows from the normalization of $\Pi(k)$, and the equation for the chemical potential, given the average density and the temperature takes the form
\begin{equation}
    I(\beta,\mu)=\frac{\rho}{1-\rho a},\label{eq:Tonksmurho}
\end{equation}
and therefore
\begin{equation}
    \Pi(k)=\frac{1-\rho a}{2\pi \rho}f(\hbar^2k^2/2m).
\end{equation}
Equation (\ref{eq:Tonksmurho}) shows that, as is well known, the Tonks-Girardeau gas with finite diameter $a$ is equivalent to a free Fermi gas, but with a rescaled density $\tilde{\rho}=\rho/(1-\rho a)$.

\paragraph{Quantum Calogero-Sutherland model.}
The quantum phase shift is again of the form $\theta(k)=-(c/2)\mathrm{sgn}(k)$, with $c$ a positive dimensionless constant. We readily obtain
\begin{equation}
    \Pi(k) = \frac{1}{2\pi (1+c/2\pi)\rho}\frac{1}{e^{\beta\left[\hbar^2k^2/2m-\tilde{\mu}(c,\beta)\right]}+1},
\end{equation}
where we have defined $\tilde{\mu}(c,\beta)=\mu+\log(1+c/2\pi)/\beta$. Normalization of $\Pi(k)$ implies
\begin{equation}
    I(\beta,\tilde{\mu}(c,\beta))=\left(1+\frac{c}{2\pi}\right)\rho.\label{eq:CS-calculation}
\end{equation}
\begin{figure}[t]
    \centering
    \includegraphics[width=\columnwidth]{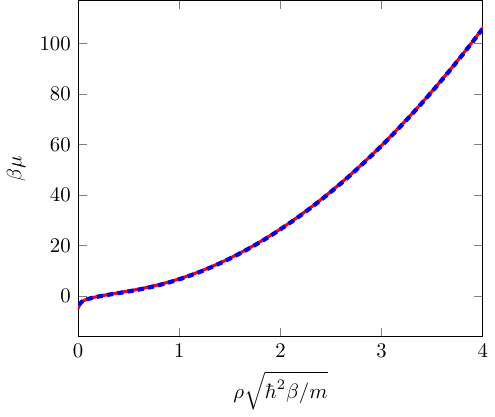}
    \caption{Chemical potential as a function of density for the quantum Calogero-Sutherland model with $c=1$. Blue dashed line: calculated using Eq.~(\ref{eq:CS-calculation}); red solid line: calculated using usual Yang-Yang thermodynamics \cite{Sutherland2004}.}
    \label{fig:Sutherland}
\end{figure}
The relation above is identical to the one for a free Fermi gas with rescaled chemical potential $\tilde{\mu}$ and density $\tilde{\rho}=(1+c/2\pi)\rho$. In fact, we have just shown that the thermodynamics of the quantum Calogero-Sutherland model is equivalent to that of a free Fermi gas but with rescaled quantities. Therefore, the thermodynamics of both the Tonks-Girardeau and Calogero-Sutherland models can be calculated directly from that of a free Fermi gas, upon rescaling of density (Tonks-Girardeau and Calogero-Sutherland) and chemical potential (Calogero-Sutherland). 

In Fig.~\ref{fig:Sutherland}, we plot the chemical potential as a function of density for the quantum Calogero-Sutherland model, using both the theory developed in this work, as well as Yang-Yang thermodynamics. Perfect agreement is obtained. 

\paragraph{Conclusions and outlook.} We have developed  methods to calculate the distributions of rapidity as well as the thermodynamic properties of integrable classical and quantum gases. Our results allow for direct computation of experimentally measurable quantities and, when they overlap, these are identical to the results obtained using standard Yang-Yang thermodynamics. The methodology here presented unveils a one-to-one correspondence between the quantum Calogero-Sutherland model and the ideal Fermi gas in thermal equilibrium. The direct nature of our results may allow for an extension to other non-standard distributions such as generalized Gibbs ensembles \cite{Rigol2008}, and the analysis of experimental results in integrable or near-integrable systems.

\paragraph{Acknowledgments.} This work was supported by the Ministry of Science, Innovation and Universities of Spain
through the Ram{\'o}n y Cajal Program (Grant No. RYC2020-029961-I), and the national
research and development grant PID2021-126039NA-I00.

\end{document}